\documentclass[twocolumn]{aastex7}
\usepackage{makecell}
\usepackage{amsmath}
\usepackage{CJKutf8}
\usepackage{textcomp}

\begin{document}

\title{What is Powering the Enigmatic He\,\textsc{II} Emitter {\it Hebe}: The First Stars or Black Holes?}

\author[0000-0002-6038-5016]{Junehyoung Jeon}
\thanks{These authors contributed equally to this work.}
\affiliation{Department of Astronomy, University of Texas, Austin, TX 78712, USA}
\affiliation{Cosmic Frontier Center, The University of Texas at Austin, Austin, TX 78712, USA}
\email[show]{junehyoungjeon@utexas.edu}

\author[0009-0000-8108-6456]{Tae Bong Jeong}
\thanks{These authors contributed equally to this work.}
\affiliation{Department of Astronomy, University of Texas, Austin, TX 78712, USA}
\affiliation{Cosmic Frontier Center, The University of Texas at Austin, Austin, TX 78712, USA}
\email{taebong.jeong@utmail.utexas.edu}

\author[0000-0003-1541-177X]{Saiyang Zhang
\begin{CJK*}{UTF8}{bsmi}(張賽暘)\end{CJK*}}
\thanks{These authors contributed equally to this work.}
\affiliation{Cosmic Frontier Center, The University of Texas at Austin, Austin, TX 78712, USA}
\affiliation{Department of Physics, University of Texas at Austin, Austin, TX 78712, USA}
\affiliation{Weinberg Institute for Theoretical Physics, University of Texas, Austin, TX 78712, USA}
\email{szhangphys@utexas.edu}

\author[0000-0003-0212-2979]{Volker Bromm}
\affiliation{Department of Astronomy, University of Texas, Austin, TX 78712, USA}
\affiliation{Cosmic Frontier Center, The University of Texas at Austin, Austin, TX 78712, USA}
\affiliation{Weinberg Institute for Theoretical Physics, University of Texas, Austin, TX 78712, USA}
\email{vbromm@astro.as.utexas.edu}


\begin{abstract}
Recent high-resolution spectroscopy with the James Webb Space Telescope (JWST) has confirmed the presence of a strong \ion{He}{2}\,$\lambda1640$ emitting clump in the vicinity of GN-z11, with only upper limits on its metallicity. To explain the peculiar properties of this source, now termed {\it Hebe}, a cluster of metal-free, Population~III (Pop~III) stars has been invoked. A less likely source for the hard UV ionizing radiation could be an accreting supermassive black hole (SMBH) embedded inside {\it Hebe}. We here provide further constraints on what could power the observed emission lines in {\it Hebe}. Comparing with cosmological simulations of Pop~III star cluster formation, we assess the maximum Pop~III stellar mass that could plausibly form at the location of {\it Hebe}, finding stellar masses of a few $10^5\,M_{\odot}$, consistent with those inferred from the observations. Modeling the continuum spectral energy distribution arising from an accreting SMBH, we derive \ion{He}{2} and \ion{H}{1} ionizing rates and the resulting recombination line luminosities, providing a less natural fit for the combined observations. We thus confirm the interpretation of {\it Hebe} as a remarkable, primordial object, with the most plausible power source provided by a massive cluster of Pop~III stars, at the limit of what is allowed within the standard model of first star formation.   
\end{abstract}


\section{Introduction} \label{sec:intro}
The pursuit of Population~III (Pop~III), the first generation of metal-free stars, is compelling \citep[e.g.,][]{Bond1981,Bromm2004,Bromm2013,Klessen2023}, given the intricate connection to cold dark matter (CDM)-driven cosmological structure formation \citep[e.g.,][]{Couchman1986,Tegmark1997}.  However, Pop~III stars have remained elusive, both at high-redshifts \citep[e.g.,][]{Schauer_ULT2020}, and during extensive, `stellar-archaeological' surveys of the Local Group \citep[e.g.,][]{Beers2005,Hartwig2015}. With the launch of the James Webb Space Telescope (JWST), the search for Pop~III has entered a new phase, focused on possible `late-forming' systems during the epoch of reionization \citep{Liu2020_2,Venditti2023,Fujimoto2025}.  

The main challenge for such late Pop~III star formation is the rapid chemical enrichment, predicted in the wake of the first supernovae, establishing a `bedrock' metallicity of $\sim 1$\% solar within the first galaxies \citep[e.g.,][]{Pallottini2014,Jaacks2018}. Early JWST surveys have confirmed this prediction, based on extensive spectroscopic follow-up \citep[e.g.,][]{Nakajima_MZR2023}, out to the highest redshifts, $z\gtrsim 12$, reached so far \citep[e.g.,][]{DEugenio2024}. Due to the highly inhomogeneous nature of early metal enrichment, driven by turbulent transport processes \citep[e.g.,][]{Ji2015,Mead2025}, pockets of pristine gas may survive until $z\sim 5$, at the tail-end of the probability distribution. The recently discovered AMORE6 galaxy at $z\simeq 5.7$ is a promising candidate, with tight upper limits on its metal content of $\lesssim 0.1$\% solar \citep{Morishita_AMORE6_2025}. 
The strongly lensing-magnified LAP1 system exhibits similar characteristics at even higher redshifts of $z\simeq 6.6$, including multiple hydrogen Lyman and Balmer line detections \citep{Vanzella_LAP12023}.
The key question is whether such pockets could host Pop~III star formation that is sufficiently massive, and therefore luminous enough, to be observable with JWST \citep[e.g.,][]{Jeong2026}. 

Due to the high photospheric temperatures predicted for Pop~III stars ($\lesssim 10^5$\,K), a key signature of metal-free star formation is the presence of strong \ion{He}{2} emission lines \citep[e.g.,][]{Tumlinson2000,BKL2001,Oh2001,Schaerer2002,Johnson_starburst2009}. With JWST, such \ion{He}{2} emitters powered by Pop~III may finally have come within reach \citep[e.g.,][]{Nakajima_diagnostics2022,Venditti2024,Venditti2026,Katz2025,Fisk2026}. Indeed, a promising candidate for a surviving Pop~III pocket was recently discovered close to the luminous GN-z11 galaxy at $z=10.6$, initially observed only through its strong \ion{He}{2} emission, with only upper limits on the continuum and any metal lines \citep{Maiolino_GNz112024}. Based on deep JWST integral-field spectroscopy, this identification has been confirmed, also with the additional detection of hydrogen H$\gamma$ emission  \citep{MaiolinoHebe2026arXiv, UblerHebe2026arXiv}.

What is powering this remarkable source, now termed {\it Hebe} by \citet{MaiolinoHebe2026arXiv}? These authors suggest a Pop~III star cluster as the most plausible explanation, using the observations to place constraints on the primordial initial mass function (IMF; \citealt{RustaHebe2026arXiv}), but also discuss accreting supermassive black holes (SMBHs) as alternatives. We here revisit the physical nature of {\it Hebe}, constraining the Pop~III stellar mass that could plausibly form in the vicinity of GN-z11, based on cosmological simulations of Pop~III star forming systems. Such independent constraint can help break the degeneracies in the analysis of \citet{RustaHebe2026arXiv}. For completeness, we also assess the plausibility of a SMBH power source inside {\it Hebe}, exploring both direct-collapse and primordial black hole (BH) scenarios.      



\section{Theoretical Models} \label{sec:models}

We investigate two main scenarios to explain {\it Hebe}: A Pop~III star cluster or an accreting SMBH, arising from a heavy-seed pathway \citep[e.g.,][]{Smith2019_2, Inayoshi2020}. Both sources could provide large amounts of hard UV-ionizing radiation to produce the observed \ion{He}{2} emission. For the flat $\Lambda$CDM cosmological parameters, we adopt a matter density of $\Omega_{\rm m} = 1 - \Omega_{\rm \Lambda}= 0.315$, baryon density $\Omega_{\rm b}=0.048$, $h = 0.6774$, and normalization parameter $\sigma_{\rm 8} = 0.829$ \citep{Planck2016}.

\subsection{Population~III Star Cluster}\label{sec:starcluster}

Previous studies have shown that a Pop~III stellar system could produce strong \ion{He}{2} line emission ($L_{\rm HeII\lambda1640}\geq 10^{40}\, \rm erg ~\, s^{-1}$) to match {\it Hebe} \citep{MaiolinoHebe2026arXiv,Venditti2026}. We therefore further investigate this scenario, under the assumption that {\it Hebe} is hosting a Pop~III star cluster.
\par
Multiple physical processes can influence the mass of the Pop~III starburst. Dominant among them is the strong radiation feedback from nearby sources, especially via photons in the Lyman-Werner (LW) band \citep[e.g.,][]{Visbal2017, Sugimura2024, Jeong2026}. For example, \citet{Jeong2026} showed a significant correlation between the Pop~III starburst mass and the strength of the local LW flux, with a maximum of $M_{\star \rm , Pop ~III} \approx 10^6 ~M_{\odot}$. This limiting stellar mass is possibly realized when the galaxy reaches a dynamical mass of $M_{\rm vir} \approx10^8~ M_{\odot}$, resulting in a strong starburst on short timescales ($\Delta t\lesssim5 \rm ~Myr$). Here, we explore GN-z11 as the source of the LW flux that is irradiating the {\it Hebe} clump, located at a measured distance of $\sim 3\rm ~pkpc$ \citep{MaiolinoHebe2026arXiv}.

\subsubsection{LW flux from GN-z11}\label{sec:lwflux}
\label{J_LW_calc} 
When estimating the LW flux from GN-z11 that reaches {\it Hebe}, we adopt an idealized analytic approach, as follows. With the distance between GN-z11 and {\it Hebe} fixed at $d \sim 3\, \rm pkpc$, we model the density of the intervening intergalactic medium (IGM) as $\rho_{\rm IGM} = \Omega_{\rm b} \rho_{c,0} (1+z)^{3}$, and assume an IGM molecular hydrogen fraction of $f_{\rm H_{2}} = 10^{-6}$ \citep{Galli2013}. Following the usual custom, we normalize the LW flux, $J_{\rm LW}$, to $J_{21}=10^{-21}\,{\rm erg\,s^{-1}\,cm^{-2}\,Hz^{-1}\,sr^{-1}}$.
We calculate the average spectral luminosity, $\langle L_{\nu}\rangle$, in the LW waveband ($h\nu = 11.2-13.6\rm eV$) using BPASS \citep{Eldridge2017}, for a simple stellar population (SSP), adopting stellar properties inferred for GN-z11: $M_{\star} = (5.6\pm0.6) \times10^8 M_{\odot}$, $t_{\rm age} = 21 \pm 3 \rm~ Myr$ \citep{Crespo-Gomez2026}.
\par
The resulting LW flux in the vicinity of GN-z11 is
\begin{equation}
\begin{split}
    J_{\rm LW, GN-z11} &= \frac{1}{4 \pi}\frac{f_{\rm esc, LW}\langle L_{\nu}\rangle}{4\pi d^2 J_{21}} \\
    &= \frac{f_{\rm esc, LW}}{(4\pi d)^2 \Delta \nu J_{21}}\int^{\nu_{\rm max}}_{\nu_{\rm min}} {L_{\nu}d\nu} \approx 1.9\times 10^3,
\end{split}
\end{equation}
where $\Delta\nu = \nu_{\rm max}-\nu_{\rm min}$, and $f_{\rm esc, LW}$ is the escape fraction of LW photons from GN-z11. We choose $f_{\rm esc, LW} = 0.64$, the conservative estimate from \citet{Schauer2017} for $\rm H$ and $\rm H_{2}$ shielding, while adopting a Kroupa IMF for the metal-enriched Population~II (Pop~II) stellar content in GN-z11.
\par
We estimate the self-shielding effect from $\rm H_{2}$ in the IGM between GN-z11 and {\it Hebe} with the prescription in \citet{Wolcoott-Green2011}: 
\begin{equation}
    \begin{split}
        f_{\rm shield}(N_{\rm H_{2}}, T) = \frac{0.965}{(1+x/b_5)^\alpha}+\frac{0.035}{(1+x)^{0.5}} \, \\
        \times \exp\left[-8.5 \times 10^{-4}(1+x)^{0.5}\right],
    \end{split}
\end{equation}
where $x \equiv N_{\rm H_{2}}/5 \times10^{14} \rm \,cm ^{-2}$, and $N_{\rm H_{2}}$ is the molecular hydrogen column density. In this equation, $b_5 \equiv b/10^5 \rm \,cm\,s^{-1}$, where $b \equiv \sqrt{k_BT/m_{\rm H}}$ is the Doppler broadening parameter, and $\alpha = 1.1$. The column density of molecular hydrogen is approximately
\begin{equation}
    N_{\rm H_{2}} = \frac{f_{\rm H_{2}} \times \rho_{\rm IGM} \times 0.752}{2 m_{\rm H}}\times d \approx10^{12}{\,\rm cm^{-2}}.
\end{equation}
Here, we assume that the IGM temperature is fixed at $T = 10^3 \rm \,K$, representing the volume-average at mean density \citep{Garaldi2022}. We have neglected the HI-shielding effect due to Lyman series absorption in the IGM \citep{Haiman2000}, as it would become significant only at $N_{\rm H} \gtrsim 10^{22}\rm \,cm^{-2}$ \citep[e.g.,][]{Neyer2022}. Therefore, the effective $J_{\rm LW}$ experienced by {\it Hebe} is 
\begin{equation}
    J_{\rm LW, eff} = f_{\rm shield}(N_{\rm H_{2}}, T) \times J_{\rm LW, GN-z11} \sim 1,800.
\end{equation}

We cross-check this estimate by investigating the merger history of the combined GN-z11 and {\it Hebe} system, using the semi-analytic model (SAM) A-SLOTH \citep{Hartwig2022,Hartwig2024,Magg2022}. The detailed star formation and LW  prescriptions can be found in previous works \citep{Liu2024,Jeon2025}. Within the SAM, we represent the global LW background with an idealized model as \citep{Greif2006,Hartwig2022}:
\begin{equation}\label{globallw}
    J_{LW,\rm global}/J_{21} = 10^{2-z/5} \mbox{\ .}
\end{equation}

To calculate the local LW flux for a given halo in the merger tree, we determine the LW photon production rate based on the total mass of active massive stars ($>5$ M$_\odot$) using the fitting formula from \citet[][equ.~8]{Deng2024}. We assume that the high-mass stars are located on average at a distance of 3\,pkpc, matching the distance between GN-z11 and {\it Hebe}. The sum of the global and local LW components is the total LW flux for the halo.

\begin{figure}[!t]
\centering
\includegraphics[width=\linewidth]{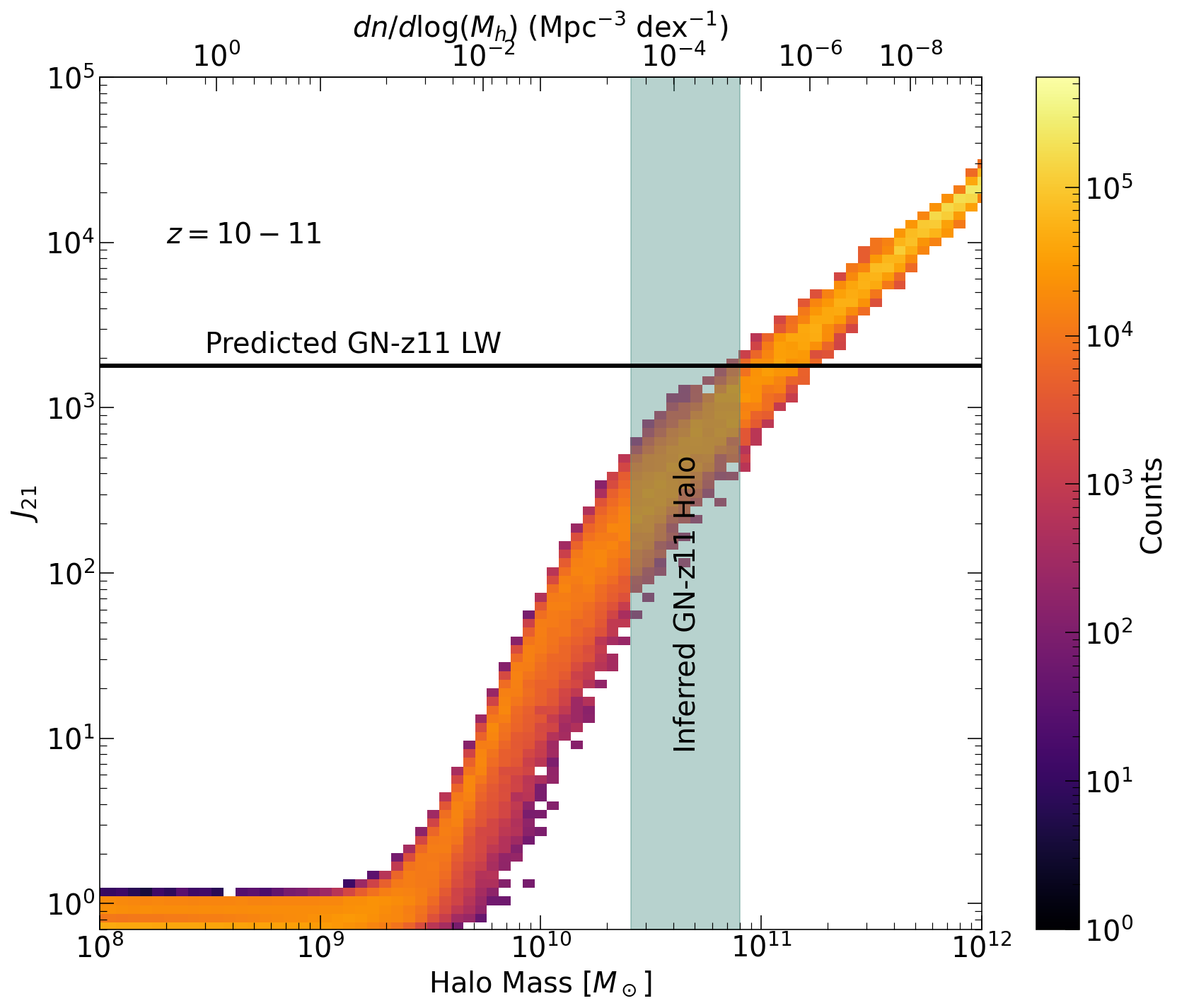}
\caption{LW fluxes within the Lagrangian volumes of biased halos at $z=9$, derived from A-SLOTH merger trees. We show the distribution drawn from the $z=10-11$ time slice, and for comparison mark the LW flux from GN-z11 as predicted above (Sec.~\ref{sec:lwflux}), together with its inferred halo mass \citep{Scholtz2024_gnz11,Tacchella2023}}. We also reproduce halo number densities from the GUREFT mass function \citep{Yung2024}, indicating that halos able to produce such high LW fluxes are rare ($\sim10^{-4}$ Mpc$^{-3}$ dex$^{-1}$), while still being within reach of current JWST surveys.
    \label{fig:lw}
\end{figure}

Fig.~\ref{fig:lw} shows the resulting distribution of halo masses vs. LW flux for 100 merger trees of halos between $2\times10^{11}$ M$_\odot-2\times10^{12}\,M_\odot$ at $z=9$. These parameters for our target halos provide an approximate representation of the biased (overmassive) environment of GN-z11. We find that halos with the inferred mass of the GN-z11 host, $\sim2-8\times10^{10}\,M_\odot$ \citep{Scholtz2024_gnz11,Tacchella2023}, can produce the LW flux as estimated above, thus further supporting our overall argument here. We note that, based on the GUREFT halo mass function \citep{Yung2024}, the number densities of such GN-z11-like systems are low ($\sim10^{-4}$ Mpc$^{-3}$ dex$^{-1}$), but still within reach of current JWST surveys (with effective survey volumes at $z\sim 10$ of $\sim10^5$\,Mpc$^3$; see, e.g., \citealt{Venditti2024}). 

\subsubsection{Pop~III Starburst Mass Prediction}\label{sec:starburst}
From the estimated effective LW flux for GN-z11 ($J_{\rm LW, eff}$) in Section \ref{J_LW_calc}, we can constrain the Pop~III starburst mass of \textit{Hebe}. Using simulated results from \citet{Jeong2026}, we derive the following fitting formula for the Pop~III starburst mass as a function of the LW flux,
\begin{equation}
    \log M_{\star,\rm Pop~III} (J_{\rm LW}) = \frac{b}{(1+ a \times \exp(-2k\log J_{\rm LW}))} + c,
\end{equation}
where $a, b, c,$ and $k$ are free parameters. We evaluate the fit for two phases, the initial starburst phase ($\langle t_{\rm age}\rangle \simeq 1.5\rm \,Myr$), and the time when the Pop~III starburst mass is maximized ($\langle t_{\rm age}\rangle \lesssim 2.5\rm \,Myr$).

\begin{figure}
    \centering
    \includegraphics[width=\linewidth]{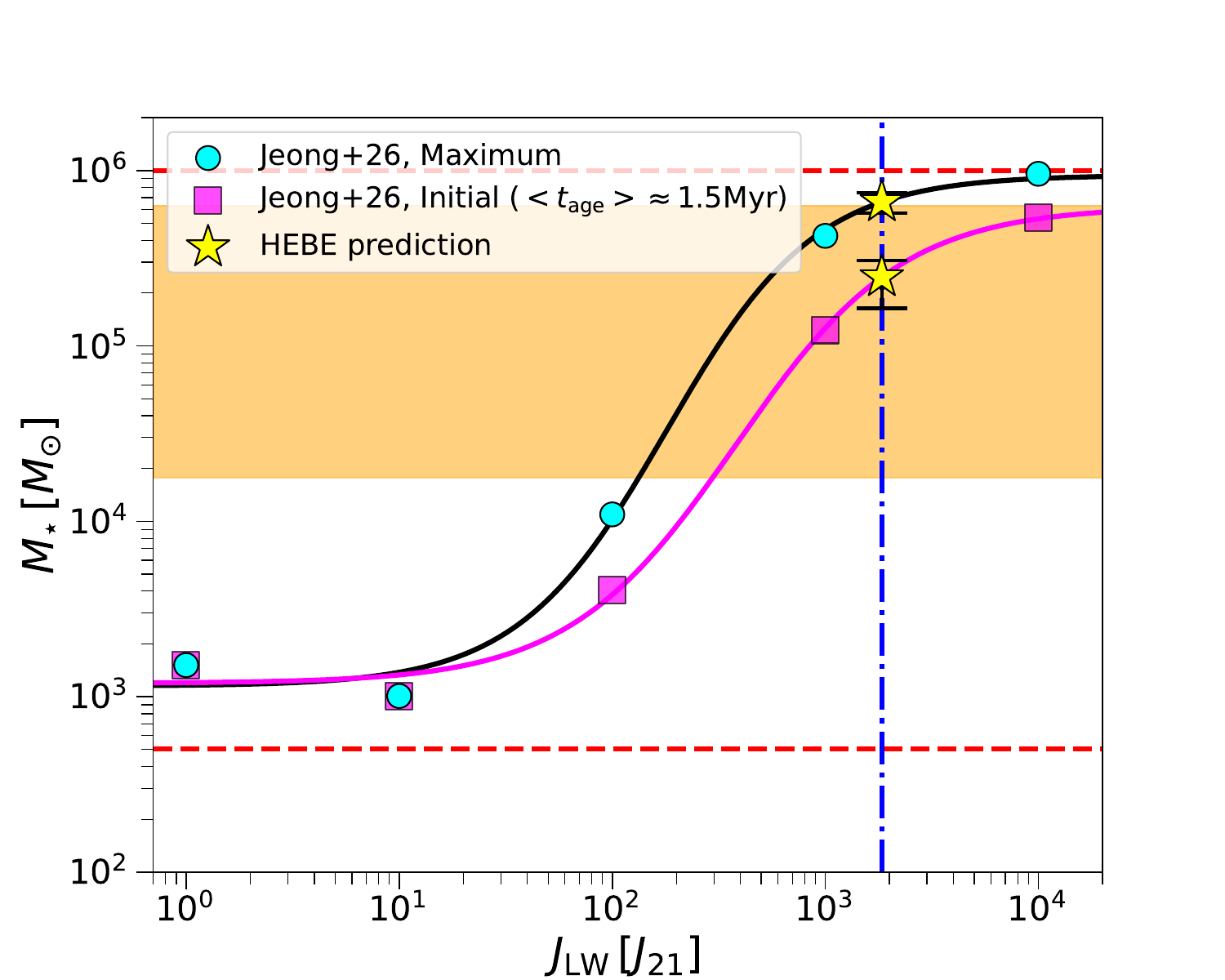}
    \caption{Pop~III starburst mass vs. strength of LW flux. We reproduce the results from the cosmological simulation in \citet{Jeong2026}, showing the initial Pop~III starburst phase ($\langle t_{\rm age}\rangle \simeq 1.5 \rm \,Myr$) with magenta squares, and the time when the maximum Pop~III mass is reached ($\langle t_{\rm age}\rangle\lesssim 2.5 \rm Myr$) with cyan circles, together with the fitting results (black and magenta solid lines). We mark the Pop~III starburst mass predicted for {\it Hebe} with yellow-star symbols (including 1$\sigma$ error bars), based on the estimated LW flux (blue dash-dotted line).
    The orange shaded region indicates the mass range for {\it Hebe}, as estimated in \citet{RustaHebe2026arXiv}.}
    \label{fig:max-pop3-mass}
\end{figure}

As can be seen in Fig.~\ref{fig:max-pop3-mass}, we infer that the Pop~III starburst mass in {\it Hebe} (yellow stars) is $M_{\rm \star, Pop~III} \sim 2.5\times 10^5~ M_{\odot}$ for the initial Pop~III starburst, and $M_{\rm \star, Pop~III} \sim 6.6\times 10^5~ M_{\odot}$ when the maximum mass is attained. These results are fully consistent with the results in \citet{RustaHebe2026arXiv}, which suggest a total stellar mass for {\it Hebe} in the range of $10^{4.25}~M_{\odot}\leq M_{\star}\leq10^{5.8}~M_{\odot}$ at $t_{\rm age} = 2\rm ~Myr$, depending on the functional form of the Pop~III IMF.

The survival of a (nearly-) pristine pocket in the vicinity of GN-z11 is clearly puzzling. Fully addressing it with self-consistent cosmological simulations is beyond the scope of this exploratory paper, but we here briefly assess whether the star formation of GN-z11 could have pre-enriched its nearby environment with metals to preclude Pop~III formation. Following the model of \citet{Jaacks2018,Jaacks2019,Liu2020}, we approximate the total supernova (SN) energy from a stellar system of mass $m_*$ as $E_{\rm SN} \simeq10^{52}~\text{erg}\times m_*/10^3 M_\odot$, 
and the final (physical) radius of SN shell expansion that enriches the environment with metals as 
\begin{equation}
    r_{\rm SN} \simeq 1~\text{kpc}\left(\frac{E_{\rm SN}}{10^{52}~\text{erg}}\right)^{0.38} .
\end{equation}
The stellar mass needed to reach $r_{\rm SN}=3$~kpc is $m_*\sim3\times10^4\, M_\odot$, much less than what is inferred for GN-z11, rendering {\it Hebe}'s lack of metal enrichment surprising, even when considering that it may have originated further away in the cosmic web, and is now in the process of merging with GN-z11. However, metal enrichment, driven by turbulent mixing processes on smaller scales, is highly inhomogeneous, and the SN blastwaves may preferentially travel into the lower-density regions of the cosmic web, thus avoiding the filaments where {\it Hebe}-like pockets may be located \citep[e.g.,][]{Ritter2012}. Again, we defer a more in-depth treatment of metal enrichment in the emerging GN-z11/{\it Hebe} system to future work.

\subsection{Supermassive Black Holes}\label{sec:smbh}

\begin{figure*}[!htb]
\gridline{
\fig{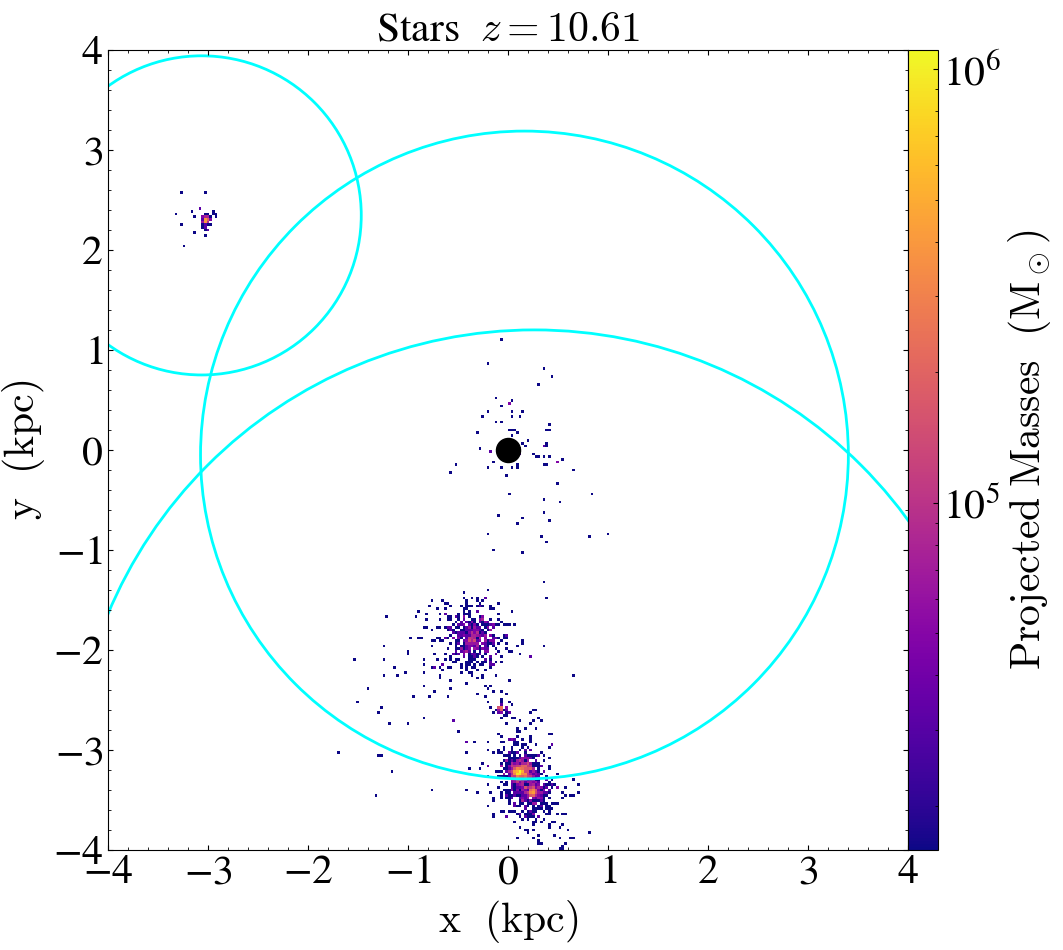}{0.5\textwidth}{}
\fig{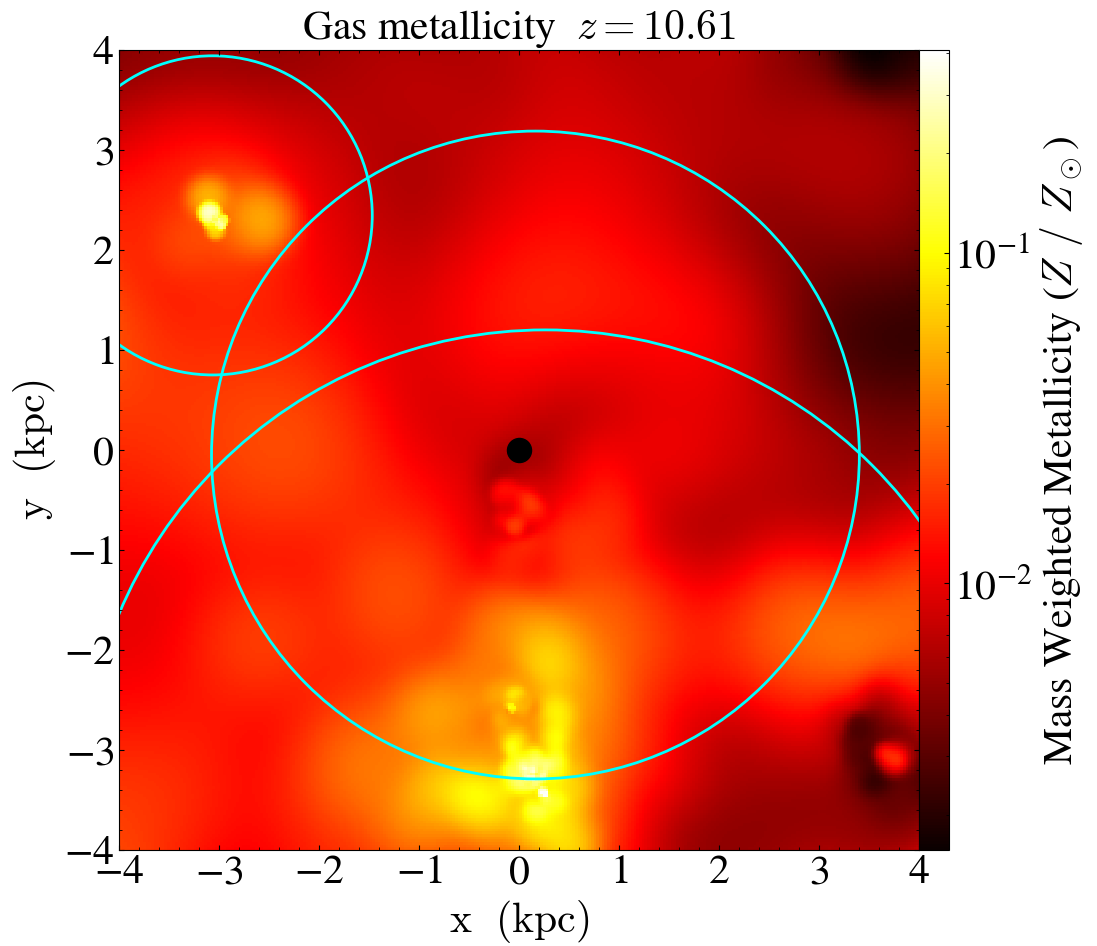}{0.51\textwidth}{}
}
\vspace{-3em}
    \caption{Example of a heavy-seed BH located close to a massive stellar dominated system, taken from the simulation suite in \citet{Jeon2024}. This configuration is analogous to GN-z11 and {\it Hebe} within the SMBH scenario.  
    We show the stellar mass (\textit{left}) and metallicity (\textit{right}) in projection, with dark matter host halos ($\sim10^8-10^9$ M$_\odot$) indicated as cyan circles. As can be seen, the main stellar component is separated by $\sim3$\,pkpc from the SMBH, marked as a black dot. The simulated SMBH has a mass of $\sim10^6$ M$_\odot$ originating from a DCBH, and experiencing some subsequent growth. The actual {\it Hebe} system may originate from a similar configuration, but with less efficient SMBH growth \citep[e.g.,][]{Jeon2023}, where a DCBH or a PBH initially forms in a metal-poor environment, and merges with a nearby metal-enriched stellar system formed separately.}
    \label{fig:bhimage}
\end{figure*}

Another scenario for {\it Hebe} is that the \ion{He}{2} emission originates from an accreting SMBH. However, any such SMBH origin is severely constrained by the observed narrow line widths, implying central BH masses of $M_\bullet\lesssim 10^4 M_\odot$ \citep{MaiolinoHebe2026arXiv}. For completeness, we here further consider the plausibility of the SMBH alternative, focusing on heavy-seed pathways\footnote{The metallicity and timing constraints for {\it Hebe} effectively exclude models where a light, stellar-remnant, seed would grow via rapid accretion.}, such as a direct-collapse black hole (DCBH) or a primordial black hole (PBH).

\subsubsection{BH Seed Formation}

The DCBH pathway is based on the runaway collapse of a massive, extremely metal-poor \citep[$Z\lesssim 10^{-3}\ \rm Z_\odot$;][]{Chon2024} gas cloud, involving a supermassive star (SMS) as an intermediate, short-lived stage \citep[e.g.,][]{Bromm2003,Begelman2006,Lodato2006}. This channel produces a more massive BH seed ($\sim10^4-10^6$ M$_\odot$; \citealt{Becerra2018b}), and requires rare conditions that suppress low-temperature gas cooling mechanisms, such as a nearby stellar population providing LW radiation that destroys molecular hydrogen, allowing the cloud to collapse without fragmenting into a large number of ordinary/low-mass stars \citep[e.g.,][]{Johnson2013,Wise2019,Haemmerl2018,Haemmerle2020}. Thus, DCBHs are initially expected to form without a host stellar population, but separate from it.

Another pathway invokes PBHs, theorized to form shortly after the Big Bang from the collapse of primordial overdensities and could span a broad range of masses, including those relevant for heavy BH seeds~\citep[$\gtrsim10^4$ M$_\odot$;][]{Zeldovich1967SvA....10..602Z, hawking1971gravitationally,Carr1975ApJ...201....1C, Belotsky2019, Escriva2022}. Unlike the DCBH channel, the PBH scenario does not rely on a pre-existing star-forming environment. Instead, PBHs may begin to influence their surroundings at very early times. In particular, efficient radiative feedback from PBH accretion can suppress gas cooling, hinder fragmentation, and delay or regulate the onset of star formation \citep[see e.g.,][]{Boyuan2022MNRAS.514.2376L, Zhang2025}. This may naturally lead to a configuration in which BH growth precedes, or is at least partially decoupled from, the assembly of a nearby stellar structure like GN-z11.

Both pathways can produce a system where the main stellar population, analogous to GN-z11, is separate from the \ion{He}{2} emitter, here assumed to be powered by an embedded SMBH. In Fig.~\ref{fig:bhimage}, we show an example of such a configuration for the DCBH pathway, reproduced from the numerical simulations in \citet{Jeon2024}. The DCBH initially forms in a metal-poor dark matter halo with low stellar mass, but soon merges with a different halo, which had formed a massive stellar population with the corresponding supernova enrichment. As metal enrichment is highly inhomogeneous in the early Universe \citep[e.g.,][]{Pallottini2014, Jaacks2019}, such disparate regions can exist in relatively close proximity. As mentioned above, fully addressing this intricate, multi-scale metal-transport problem requires dedicated follow-up simulations, and is beyond the scope of our exploratory study here.

\subsubsection{SMBH SED Modeling}\label{sec:SED}

\begin{figure*}[t]
    \centering
    \includegraphics[width=0.95\textwidth]{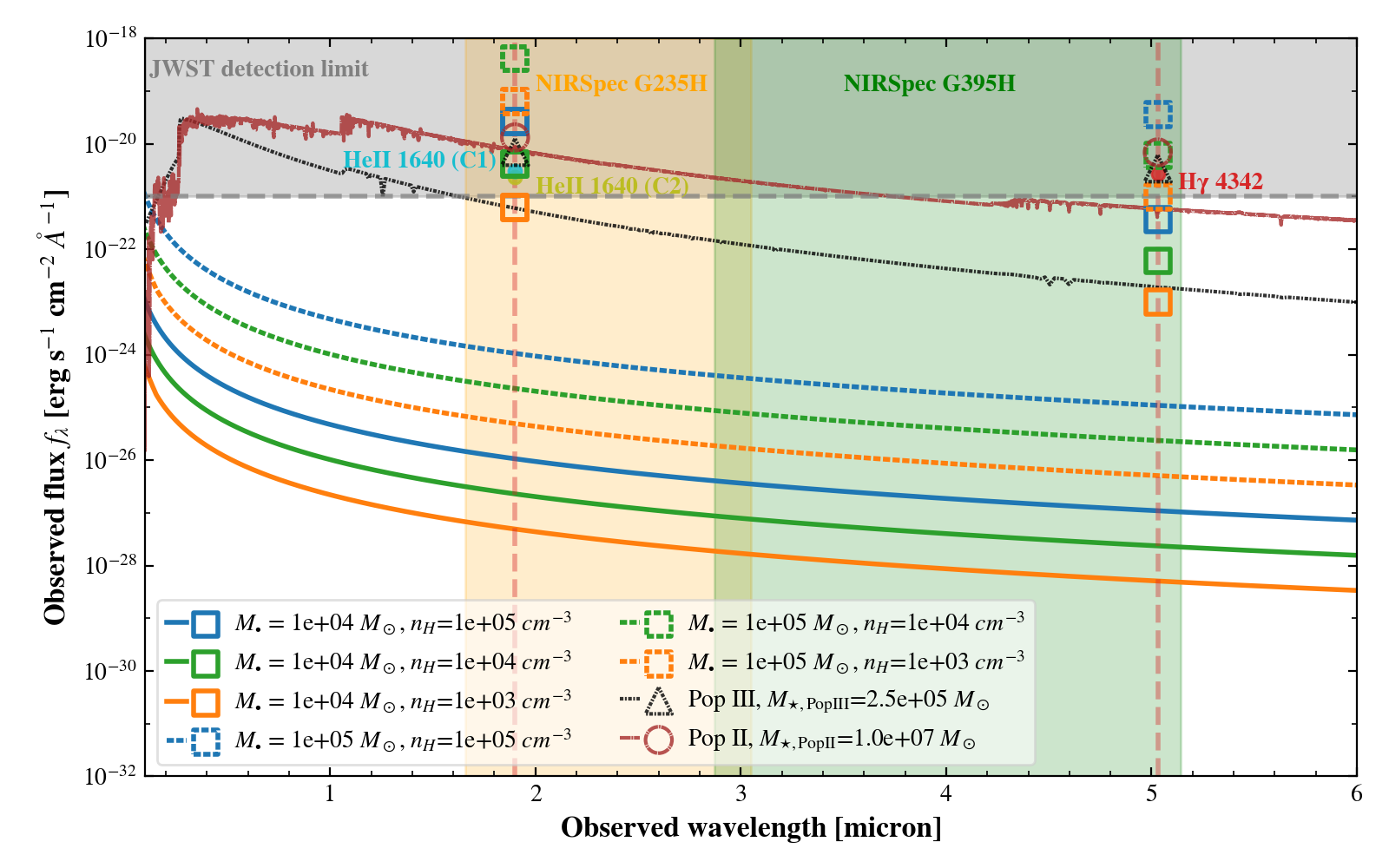}
\caption{Observer-frame spectral energy distribution of a BH accreting at the Bondi rate within a halo of mass $M_h \sim 10^8\,M_\odot$ at $z=10.6$, for BH masses $M_\bullet=10^4$ and $10^5\,M_\odot$ (solid and dotted lines, respectively), and ambient densities $n_{\rm H}=10^3$--$10^5\,\mathrm{cm}^{-3}$ (colors, as described in the legend). The BH rest-frame SED is modeled using the prescription from~\citet{Takhistov2022JCAP}.
For comparison, we also show a Pop~III and Pop~II simple stellar population with total stellar mass $M_{\star,\mathrm{PopIII}}=2.5\times10^5\,M_\odot$ (black dot-dashed line) and $M_{\star,\mathrm{PopII}}=1.0\times10^7\,M_\odot$ (brown dot-dashed line), respectively, computed using the Yggdrasil spectral synthesis model~\citep{Zackrisson2011} and BPASS~\citep{Eldridge2017} at an ambient density of $n_{\rm H}=10^3\,\mathrm{cm}^{-3}$. Filled circles mark the observed line-equivalent flux densities $f_\lambda$ of \ion{He}{2}~$\lambda1640$ (C1 and C2 components) and H$\gamma$~$\lambda4342$ reported by~\citet{UblerHebe2026arXiv, MaiolinoHebe2026arXiv}. Open squares, triangles and circles of matching color and line style show the predicted reprocessed line flux densities obtained by integrating the ionizing photon rate above the relevant threshold, assuming complete photon reprocessing ($f_{\mathrm{esc}}=0$). The orange and green shaded bands indicate the wavelength coverage of the JWST/NIRSpec G235H and G395H gratings, and the gray shaded band indicate the JWST flux density detection threshold. The vertical dashed lines mark the redshifted wavelengths of the \ion{He}{2}~$\lambda1640$ and H$\gamma$~$\lambda4342$ transitions. We have verified that varying the density over
 $n_{\rm H} = 10$--$10^5\,\mathrm{cm}^{-3}$ does not significantly affect the predicted line flux densities. The Pop~II stellar mass of $M_{\star,\mathrm{PopII}} =1.0\times10^7\,M_\odot$ is chosen to reproduce the observed emission-line strengths, but such a population would produce continuum emission in excess of current JWST limits.}
    \label{fig:redshifted_bh_sed}
\end{figure*}

Regardless of its specific origin, we next assess whether an accreting SMBH could power the line emission observed in {\it Hebe}. We model the BH spectral energy distribution (SED) based on the prescription in \citet{Takhistov2022JCAP}, and derive the corresponding emission-line strengths. 

Given a BH of mass $M_\bullet$ embedded in gas of hydrogen number density $n_{\rm H}$, we estimate the accretion rate using a Bondi-like prescription,
\begin{equation}
    \dot{M}_{\bullet} = \frac{4\pi\,(G M_{\bullet})^{2}\,\rho_{\rm gas}}{\tilde{v}^{3}}\,,
\end{equation}
where $\rho_{\rm gas}=\mu\,m_p\,n_{\rm H}$ is the local gas mass density with mean molecular weight $\mu=1.22$ appropriate for neutral primordial gas, and $\tilde{v}$ is the effective relative velocity between the BH and the surrounding medium. We take $\tilde{v}$ to be of order the virial velocity in an atomic cooling halo,
\begin{equation}
    \tilde{v} \simeq \sqrt{\frac{G M_{\rm h}}{R_{\rm vir}}} \simeq 18.1\;\mathrm{km\,s^{-1}}\left(\frac{M_{\rm h}}{10^{8}\,M_{\odot}}\right)^{1/3}\left(\frac{1+z}{11}\right)^{1/2}\,.\label{eq:vtilde}
\end{equation}
We note that this is the dynamical mass scale inferred for {\it Hebe} \citep{MaiolinoHebe2026arXiv}.
It is convenient to express the accretion rate in dimensionless form as $\dot{m}\equiv\dot{M}_{\bullet}/\dot{M}_{\rm Edd}$, where the Eddington accretion rate is $\dot{M}_{\rm Edd}=L_{\rm Edd}/(\epsilon_r\,c^2)$ and $\epsilon_r$ the radiative efficiency. Scaling to fiducial values gives
\begin{equation}
    \dot{m} \simeq 0.25\left(\frac{\epsilon_r}{0.057}\right)\left(\frac{M_{\bullet}}{10^{4}\,M_{\odot}}\right)\left(\frac{n_{\rm H}}{10^{3}\,\mathrm{cm^{-3}}}\right)\left(\frac{\tilde{v}}{20\,\mathrm{km\,s^{-1}}}\right)^{-3}\,.\label{eq:mdot}
\end{equation}

The fiducial radiative efficiency $\epsilon_r=0.057$ corresponds to the standard optically thick efficiency for a Schwarzschild BH. The dimensionless accretion rate $\dot{m}$, compared to the critical threshold $\dot{m}_{\rm crit}\simeq 0.07\,\alpha$, determines the structure of the accretion flow and thus the shape of the emitted spectrum, where $\alpha$ is the viscosity parameter and is fixed to $\alpha=0.1$ throughout this work. For $\dot{m}>\dot{m}_{\rm crit}$, the accretion flow is assumed to be geometrically thin and optically thick, and we compute the rest-frame spectral luminosity $L_E(E)$ [$\mathrm{erg\,s^{-1}\,eV^{-1}}$] using a standard multi-color disk model \citep{Shakura1973A&A....24..337S, Pringle1981ARA&A..19..137P}. For $\dot{m}\leq\dot{m}_{\rm crit}$, the flow becomes radiatively inefficient and advection-dominated, and we instead adopt an advection dominated accretion flow (ADAF)-like prescription \citep{Narayan1995ApJ...452..710N}. In summary,
\begin{equation}
    L_E(E)=
    \begin{cases}
        L_E^{\rm thin}(E), & \dot{m}>\dot{m}_{\rm crit},\\[6pt]
        L_E^{\rm ADAF}(E), & \dot{m}\leq\dot{m}_{\rm crit}.
    \end{cases}
    \label{eq:SED}
\end{equation}
In each case, $L_E(E)$ is normalized such that integration over energy recovers the bolometric luminosity $\int L_E(E)\,dE  = \epsilon_r \dot{M}_{\bullet} c^2 \, .$

For the low-accretion branch, the subgrid model produces a harder and more extended ionizing spectrum, intended to mimic the emission from a hot, optically thin flow. By contrast, in the thin-disk branch, the spectrum is approximated as a multicolor blackbody with an exponential high-energy cutoff, so that the emission is concentrated around the characteristic inner-disk temperature. For the parameter range relevant to {\it Hebe}, this branch is particularly important for setting the number of photons above the \ion{H}{1} and \ion{He}{2} ionization thresholds, and therefore for determining the strength of the associated recombination lines.

Once the continuum SED is specified, the rate of ionizing photons above a threshold energy $E_0$ follows directly from
\begin{equation}
Q(>E_0)=\int_{E_0}^{\infty}\frac{L_E}{E}\,dE \, ,
\end{equation}
where $E_0=13.6\;\mathrm{eV}$ for hydrogen and $54.4\;\mathrm{eV}$ for singly ionized helium. In a radiation-bounded nebula where all ionizing photons are reprocessed, the luminosity of a recombination line at frequency $\nu_{\mathrm{line}}$ is
\begin{equation}
L_{\mathrm{line}}=(1-f_{\mathrm{esc}})\,\epsilon_{\mathrm{line}}\,h\nu_{\mathrm{line}}\,Q \, ,
\end{equation}
where $f_{\mathrm{esc}}$ is the escape fraction and $\epsilon_{\mathrm{line}}\equiv\alpha_{\mathrm{line}}^{\mathrm{eff}}/\alpha_B$ is the fraction of recombinations that produce the line of interest \citep{Osterbrock2006}. Throughout this work we assume $f_{\mathrm{esc}}=0$ for both hydrogen- and helium-ionizing photons, i.e., complete reprocessing within the nebula. For H$\beta\;\lambda4861$, at electron temperatures $T_e\sim1$--$3\times10^4\;\mathrm{K}$ and densities below the collisional de-excitation limit, we adopt $\epsilon_{\mathrm{H}\beta}\approx0.12$; higher-order Balmer lines then follow from the intrinsic decrement, and we take $L_{\mathrm{H}\gamma}/L_{\mathrm{H}\beta}=0.468$, appropriate for $T_e\approx2\times10^4\;\mathrm{K}$. For \ion{He}{2}~$\lambda1640$, the relevant photon rate is $Q_{\mathrm{He}^+}$, with an effective emission probability $\epsilon_{\rm He\,II\,1640}\approx0.47$ at $T_e\sim3\times10^4\;\mathrm{K}$ \citep[recombination coefficients are evaluated following][]{Storey1995MNRAS,PyNeb}. Since the $54.4\;\mathrm{eV}$ threshold lies well above the hydrogen ionization edge, the ratio $Q_{\mathrm{He}^+}/Q_{\mathrm{H}}$ is sensitive to the hardness of the ionizing spectrum and therefore to the accretion state of the BH. 

The resulting observer-frame fluxes are shown in Fig.~\ref{fig:redshifted_bh_sed}, where we explore BH masses of $10^4$ and $10^5\,M_\odot$ and ambient gas densities in the range $n_{\rm H}=10^3$--$10^5\,\mathrm{cm}^{-3}$. The corresponding accretion rates\footnote{The resulting BH rest-frame SEDs all follow the thin-disk branch in Equ.~\ref{eq:SED}.} span $\dot{m}\sim0.33$--$330$, covering the range from moderately sub-Eddington to hyper-Eddington. The predicted continuum level depends sensitively on both BH mass and gas density, with an overall spread of nearly four orders of magnitude, implying that observational limits on the continuum already provide a meaningful constraint on the underlying BH--gas configuration. The resulting line and continuum levels are broadly consistent with the observations reported in \citet{UblerHebe2026arXiv}. If part of the continuum originates from stars, models with a more modest BH-powered continuum, such as $M_{\rm BH}\sim10^4\,M_\odot$ and $n_{\rm H}\sim10^3$--$10^4\,\mathrm{cm}^{-3}$, are preferred because they leave room for a non-negligible stellar contribution. By contrast, models with larger $M_{\rm BH}$ and/or higher $n_{\rm H}$ tend to produce an overly dominant BH continuum. For completeness, we also show the continuum and corresponding emission-line strengths for a Pop~III star cluster with the mass estimated in Section~\ref{sec:lwflux}, as well as a young Pop~II stellar population ($t_{\rm age} = 3 \,{\rm Myr}, \, Z_{\star} = 0.02\, Z_{\odot}$) whose mass is chosen to best reproduce the observed line fluxes, adopting simple modeling from \citet[][equ. 3]{Raiter2010}. The Pop~III model yields predicted line strengths broadly consistent with the observations while remaining below the continuum detection threshold; the Pop~II model, however, requires a stellar mass large enough that its continuum would already be detectable by JWST. Our analysis is limited to the instantaneous configuration inferred from the data; a full reconstruction of the formation history of {\it Hebe} will require future high-resolution simulations.

With the same SED model, we also calculate the detectability of an accreting BH via X-ray emission. Only the most extreme models ($M_\bullet = 10^5\,M_\odot$, $n_H \geq 10^4$ cm$^{-3}$) produce fluxes above the Chandra Deep Field-South (CDF-S) 7 Ms limit ($\sim 1.9\times10^{-17}$ erg s$^{-1}$ cm$^{-2}$ within $0.5-7$~keV; \citealt{ChandraLimit2017}); the rest of the parameter space falls orders of magnitude below current X-ray sensitivity. This is not unique to \textit{Hebe}: the broader population of JWST-discovered AGN at high redshift appears to be systematically X-ray weak \citep{Maiolino2024}, and even GN-z11, as a spectroscopically confirmed AGN at the same redshift and in the same field, remains X-ray undetected \citep{Maiolino2023}. For comparison, the case of UHZ-1 at $z\approx10.1$, initially reported as a $\sim 4\times10^7\,M_\odot$ DCBH candidate on the basis of a Chandra hard-band detection \citep{Bogdan:2023UHZ1, Natarajan2023}, further illustrates these difficulties: a recent reanalysis of the full 2.2~Ms dataset finds the X-ray excess at only $2.3$--$2.9\sigma$ significance, while JWST/MIRI non-detections independently favor a star-forming galaxy interpretation \citep{Zou2026}. These examples underscore that X-ray observations at cosmic dawn alone carry little power to distinguish between accreting BH and stellar scenarios \citep[see also][]{Jeon2022}. X-ray non-detection of \textit{Hebe} is therefore consistent with, but not constraining for, most BH configurations considered here.


\section{Summary and Conclusions} \label{sec:conclusions}

In this work, we have examined two plausible scenarios for {\it Hebe}: a Pop~III stellar cluster or an accreting SMBH. For the Pop~III scenario, we argue that GN-z11 provides the LW flux of sufficient strength to trigger a Pop~III starburst at the location of {\it Hebe}, with an estimated mass of $\sim2.5-6.6\times10^5$ M$_\odot$, in agreement with previous work \citep{RustaHebe2026arXiv}.  
For the scenario where {\it Hebe} is an accreting SMBH ($M_{\bullet}=10^4-10^5$ M$_\odot$), originating from a DCBH or PBH, we model the emitted continuum SED, finding that the observed {\it Hebe} \ion{He}{2} and H$\gamma$ fluxes could be individually achieved for plausible choices of SMBH mass and nearby gas density. The SMBH scenario, however, provides a less natural fit to the combined flux measurements. One additional scenario involving Pop~II stellar clusters has also been tested. However, the observed line fluxes require a cluster mass of $M_{\star} \approx10^{7}M_{\odot}$, producing a detectable stellar continuum signature.

We note that these scenarios are not mutually exclusive. The high LW flux from GN-z11 could promote both a Pop~III starburst and DCBH formation \citep{Aykutalp2020,Jeon2024,Jeong2026,Fisk2026}. If the SMBH mass or gas density is low enough, a combination of stellar and SMBH contributions can produce the observed {\it Hebe} fluxes (see Fig.~\ref{fig:redshifted_bh_sed}). In any case, {\it Hebe} is a remarkable object, plausibly originating from the first stars, an exotic SMBH, formed from a DCBH or PBH pathway, or even a combination of both. 


Observing the signatures of the first stars and SMBH seeding pathways has been a longstanding objective in astronomy, and JWST's discovery of {\it Hebe} marks an important milestone. 
Within the LW-mediated model explored here, there may be a connection to the recently discovered class of `synchronized pairs' \citep[][]{Visbal2014}, in this case UV-luminous galaxies close to Little Red Dots (LRDs), the latter assumed to host DCBH-seeded SMBHs \citep[e.g.,][]{Baggen2026,Jeon2025_lrd,Cenci2025,Pacucci2026}. Future theoretical studies are needed to further unravel the bifurcation behavior of pristine gas clouds subject to strong LW fluxes, resulting in a massive Pop~III star cluster or heavy BH seed, and future observations may provide a fuller picture of how rare {\it Hebe}-like objects truly are. We are entering a new, exciting  phase in the pursuit of Pop~III stars, and the understanding of the primordial Universe that allows them to form.  

\section*{Acknowledgments}
 The authors acknowledge the Texas Advanced Computing Center (TACC) for providing HPC resources under allocation AST23026. VB acknowledges support from the Josey Centennial Professorship in Astronomy at UT Austin.

\bibliography{ms}{}
\bibliographystyle{aasjournal}

\end{document}